\documentstyle[12pt,openbib]{article}
\def\Ao{A_0^{\ \Lambda}} 
\def\Ala{A_a^{\ \Lambda}} 
\def\Ama{A_a^{\ \Gamma}} 
\def\Ela{E^a_{\ \Lambda}} 
\def\Ena{E^a_{\ \Delta}} 
\def\Alb{A_b^{\ \Lambda}} 
\def\Elc{E^c_{\ \Lambda}} 
\def\eela{E_a^{\ \Lambda}} 
\def\eelc{E_c^{\ \Lambda}} 
\def\Fl{F^{ab}_{\ \Lambda}} 
\def\ffl{F_{ab}^{\ \Lambda}} 
\def\D{{\cal D}} 
\def\J{{\cal J}} 
\def\h{h_{ab}} 
\def\p{\pi^{ab}} 
\def\H{h^{ab}} 
\def\P{\pi_{ab}} 
\def\conf{ (\h,\p,\Ala,\Ela )} 

\begin{document}
\title{ Mass Formulas for Stationary Einstein-Yang-Mills  Black Holes 
   and a Simple Proof of Two Staticity Theorems }

\author{{\normalsize Daniel Sudarsky } \\
        {\normalsize PUCMM, Santo Domingo, Dominican Republic, Apartado
2748}\\
         and\\
        {\normalsize Robert M. Wald} \\
  {\normalsize Enrico Fermi Institute, University of Chicago, 5640 S. Ellis
  Ave, Chicago, IL 60637} }
\date{} 
\maketitle 
\begin{abstract}
We derive two new integral mass 
formulas for 
stationary black 
holes in Einstein-Yang-Mills theory. From these we 
derive a 
formula for 
$ \J \Omega -Q V  $, from which 
it follows immediately that any stationary, nonrotating, uncharged
black hole 
is static and has vanishing 
electric field on the static slices. In the Einstein-Maxwell case, 
we have, in addition, the ``generalized Smarr mass formula", 
for which we provide a new, simple derivation. When combined 
with the other two formulas, we obtain a simple proof that 
nonrotating Einstein-Maxwell black holes must be static 
and have vanishing magnetic field on the static slices. 
Our mass formulas 
also can be generalized to cases with 
other types of matter fields, and we describe the nature of these 
generalizations.
\end{abstract}

\medskip 
In a recent paper by the authors \cite{Us}, it was shown, 
among other things, that 
a solution of the Einstein-Yang-Mills (EYM) equations 
describing a stationary black 
hole with  bifurcate Killing horizon and satisfying 
$ VQ = \J \Omega =0  $ is necessarily 
static and has vanishing electric field on the static slices. 
A stronger result was obtained in the Einstein-Maxwell case: 
It was proven that a solution of the Einstein-Maxwell equations 
describing a stationary black 
hole with  bifurcate Killing horizon and satisfying 
$\J \Omega =0  $ is necessarily 
static and has vanishing magnetic field on the static slices. 
These staticity theorems were obtained by 
deriving a generalized first law of 
black hole mechanics, using it to infer extremal properties 
of stationary black hole solutions, and then showing that these 
extremal properties could be violated unless the black hole is static. 
The theorems do not require the stationary Killing field to be
globally timelike in the exterior region, i.e., ``ergoregions" are permitted.
Thus, in particular, the Einstein-Maxwell staticity theorem
closed a gap in the black hole
uniqueness theorems which had been open for nearly two decades
(see \cite{Hawk1}).

The purpose of this 
paper is to derive some new  ``mass formulas" relating the 
asymptotically defined 
attributes of a stationary black hole in 
EYM theory and to use them to give 
a simple proof of the above staticity theorems. For definiteness, 
we will restrict our considerations to SU(2)-EYM theory, 
although all of our equations and results 
also apply straightforwardly 
to Einstein-Maxwell theory. More generally, mass formulas 
analogous to the ones we derive will exist for many other theories, 
and we will explain the conditions under which such generalizations 
can be obtained. 

We first briefly review the ``3 + 1" formulation of
the EYM equations given in \cite{Us}. 
Initial data in 
EYM theory consists of the 
specification of the fields $\conf $ on a 
three-dimensional manifold, $\Sigma$. 
Here $\h$ is a Riemannian metric on $\Sigma$, 
$\Ala $   
is the gauge field component tangent to $\Sigma$, 
$\p $ is the canonically conjugate momentum to $\h$, and 
$\Ela$ is 
the electric field (viewed as a density of weight 1/2), 
which is (up to the numerical factor of $1/4$) the 
momentum canonically conjugate to $\Ala$. 
Here and throughout this paper, lower case greek indices 
are used to denote 
spacetime tensors, and latin indices are used to denote 
tensors in the hypersurface $\Sigma$. The projection of a spacetime 
tensor into $\Sigma$ is denoted by replacing the greek indices 
by latin indices, e.g., the projection of the spacetime vector $t^\mu$ 
into $\Sigma$ is denoted as $t^a$. Capital greek indices are 
used for the Lie algebra of the Yang-Mills field. 

Constraints are present in Einstein-Yang-Mills theory. 
On a hypersurface, $\Sigma$, on which $\pi_a^{\ a} =0 $, 
the allowed initial data is restricted to 
those that at each point $x \in \Sigma $ satisfy 
\begin{equation}
0=\sqrt h \D_a (\Ela /\sqrt h ) =
\sqrt h D_a(\Ela/\sqrt h) + c_{\Lambda \Gamma}^{\ \ \Delta} \Ama \Ena
\label{(2.11)}
\end{equation}
\begin{equation}
0  =  -R + (1/h)\P\p  + (2/ h) \eela \Ela +   \ffl \Fl
\label{(2.12)}
\end{equation}
\begin{equation}
0=\sqrt h D_b (\pi_a^{\ b} /\sqrt h)
-2\ffl E^b_{\ \Lambda} 
\label{(2.13)}
\end{equation}
where $D_a$ is the derivative operator on $\Sigma$ compatible with 
the metric, $\h$ , $\D_a$  denotes the 
(metric compatible) gauge covariant 
derivative operator, and $R$ denotes 
the scalar curvature of $\h$. 

We shall be concerned in this paper with spacetimes representing a 
stationary black hole with a bifurcate Killing horizon. 
As discussed in \cite{Racz}, this should encompass 
all stationary black hole solutions in EYM theory 
except for the ``degenerate" 
solutions which have vanishing surface gravity. 
Recall that a stationary 
black hole with bifurcate Killing horizon automatically possesses 
a Killing field, $t^\mu$, which 
approaches a time translation 
in the asymptotic region and a Killing field, 
$\chi^\mu$, which vanishes on the bifurcation surface $S$. 
If $\chi^\mu$ fails to coincide with 
$t^\mu$, then the spacetime also posesses an axial 
Killing field $\phi^\mu$ such that 
\begin{equation}
\chi^\mu = t^\mu + \Omega \phi^\mu 
\label{(2.45)}
\end{equation}
where the constant 
$\Omega $ is known as the angular velocity of the horizon. 
It has recently been proven \cite{WaldC} that any stationary 
black hole with bifurcate Killing horizon admits 
an asymptotically flat maximal ($\pi_a^{\ a} =0 $) hypersurface 
which is asymptotically orthogonal 
to $t^\mu $ and whose boundary is 
the bifurcation surface, $S$, of the horizon. 
We choose $\Sigma$ to be such a hypersurface. 

Now, consider the evolution equations for the initial data which 
is induced on $\Sigma$. 
Choose the lapse and shift functions, $N^\mu = (N, N^a)$, 
to coincide with a Killing field in the spacetime. 
Then, the EYM evolution equations yield 
the following relations, obtained by setting $\pi_a^{\ a} =0 $ in 
the equations given in \cite{Us}: 
\begin{equation} 
0=\dot\p   =- (\sqrt h N a^{ab} + \sqrt h [\H D^cD_c (N)- D^a D^b (N)] 
-   \mbox{\rm{\pounds}}_{N^i} \p) 
\label{(2.16)} 
\end{equation} 
n\begin{equation} 
0=\dot\h = {N \over {\sqrt h }}2\P  + 
\mbox{\rm{\pounds}}_{N^i} \h 
\label{(2.17)} 
\end{equation} 
\begin{equation} 
0=\dot\Ela  = -  
(\sqrt{h} \D_b (N \Fl ) +N c_{\Lambda\Gamma}^{\ \ \Delta} 
A_0^{\ \Gamma} E^a_{\ \Delta} -\mbox{\rm{\pounds}}_{N^i} \Ela ) 
\label{(2.18)} 
\end{equation} 
\begin{equation} 
0=\dot\Ala =  
( N \eela /\sqrt{h} + \D_a (N\Ao )+ 
\mbox{\rm{\pounds}}_{N^i} \Ala) 
\label{(2.19)} 
\end{equation} 
with 
\begin{eqnarray} 
& a^{ab}  \equiv 
 (2/ h)(\Ela E^{b \Lambda} -(1/2) \H \eelc \Elc ) 
+2 (F^{ac}_{\ \Lambda} F_c^{\ b\Lambda} + 
(1/4) \H  F^{cd}_{\ \Lambda} F_{cd}^{\ \Lambda}) \nonumber \\ 
&  + (R^{ab} -(1/2) \H R) 
+(1/ h)(2\pi^a_{\ c} \pi^{bc}  
 -(1/2)\H \pi^{cd} \pi_{cd} ) \nonumber \\ 
& 
\label{(2.16b)} 
\end{eqnarray} 
where 
\begin{equation} 
\mbox{\rm{\pounds}}_{N^i} \h =2D_{(a} N_{b)} 
 \label{Lieh} 
\end{equation} 
\begin{equation} 
\mbox{\rm{\pounds}}_{N^i} \Ala =N^bD_b \Ala + \Alb D_a N^b 
\label{LieA} 
\end{equation} 
and 
\begin{eqnarray} 
\mbox{\rm{\pounds}}_{N^i} \p = \sqrt h N^c D_c (\p/\sqrt h) 
-2\pi^{c(a} D_c N^{b)} +\p D_c N^c 
\label{Liea} 
\end{eqnarray} 
\begin{eqnarray} 
\mbox{\rm{\pounds}}_{N^i} \Ela = \sqrt h N^c D_c (\Ela /\sqrt h) 
-\Elc D_c N^a +\Ela D_c N^c 
\label{Lieb} 
\end{eqnarray} 

We begin by deriving a simple equation satisfied 
by the lapse function $N = - k^\mu n_\mu$ 
associated with any Killing field $k^\mu$, where $n^\mu$ 
denotes the unit normal to the maximal hypersurface $\Sigma$. 
Contracting (\ref{(2.16)}) with $h^{ab}$, we find 
\begin{equation} 
 D_c D^c (N) = -(1/2) N a^b_{\  b} - D_a (N_b) \pi^{ab} /\sqrt h 
\label {d1} 
\end{equation} 
{}From eqs.(\ref{(2.17)}) and (\ref{Lieh}) we find 
\begin{equation} 
D_a (N_b) \pi^{ab} = -N \pi_{ab} \pi^{ab} /\sqrt h  
\end{equation} 
Substituting in (\ref{d1}) and using the 
constraint (\ref{(2.12)}) in the 
expression for $a^b_{\  b}$ we obtain 
\begin{equation} 
 D_c D^c (N) = \rho N 
\label{d3} 
\end{equation} 
where 
\begin{equation} 
 \rho = (1/h) \P\p 
  + (1/ h)  \eela \Ela + (1/2) \ffl \Fl 
\label{rho} 
\end{equation} 
so that $\rho$ is non-negative. 
Note that the derivation of eq.(\ref{d3}) used only the ``Einstein 
portion" of the EYM equations, and, thus, it is easily 
generalized to any other 
Einstein-matter system (even if the full system is not derivable from 
a Hamiltonian). Indeed, a generalization of our derivation
shows that in any spacetime foliated by 
maximal hypersurfaces, 
the lapse function, $N$, of this foliation 
satisfies eq.(\ref{d3}) with $\rho$ replaced by, 
\begin{equation} 
 \rho = (1/h) \P\p 
  +  R_{\mu \nu} n^\mu n^\nu 
\label{rho'} 
\end{equation} 
(This result also could be derived from the Raychaudhuri equation
for non-geodesic timelike congruences; see eq.(4.26) of \cite{Hawk1}.)
Thus, in particular, for a stationary black hole with 
bifurcate horizon in any Einstein-matter system, eq.(\ref{d3}) 
holds with $\rho$ given by (\ref{rho'}). Note that $\rho$ 
will be non-negative 
provided only that the matter satisfies the strong 
energy condition. When $\rho$ is non-negative, 
the maximum principle can be usefully applied to eq.(\ref{d3}), and
solutions to eq.(\ref{d3}) 
 are uniquely determined 
by their boundary value at $S$ and their asymptotic
value at infinity. 

As our first application of eq.(\ref{d3}), we choose 
$N^\mu = \phi^\mu$, where $\phi^\mu$ is the axial Killing 
field, so $N = - n_\mu \phi^\mu$. The boundary conditions 
( $N_{|S}=0 $ and $N_{|\infty } =0 $) 
yield the unique solution $N =0 $ on $\Sigma $. 
Thus, we find that $\phi^\mu$ 
is tangent to $\Sigma $. This result also could be proven by a 
generalization of known uniqueness results on maximal foliations 
(see theorem 5.5 of \cite{BCO}), since if $\phi^\mu$ failed to be 
everywhere tangent to $\Sigma$, we could obtain a new maximal 
hypersurface asymptotic to $\Sigma$ by applying a 
rotation to $\Sigma$. 

Next we apply eq.(\ref{d3}) to the stationary Killing field $t^\mu$ 
We write $\lambda$ for the lapse function, $N$, in this case 
i.e., we define, 
\begin{equation} 
 \lambda = - n_\mu t^\mu
\label{lambda} 
\end{equation} 
Since $\lambda$ satisfies the boundary conditions 
$\lambda_{|S}=0 $ and $\lambda_{|\infty } = 1 $ 
the maximum principle implies that $\lambda$ is strictly positive 
on $\Sigma$ outside of $S$. 
(Also $\lambda < 1$ throughout $\Sigma$.) 
Integrating eq.(\ref{d3}) over $\Sigma$, we obtain, 
\begin{equation} 
\int_{\infty} d S^a D_a \lambda - 
\int_{S} d S^a D_a \lambda = \int_{\Sigma} \lambda \rho 
\end{equation} 
where, here and below, all volume integrals over $\Sigma$
are taken with respect to the natural volume element determined
by $h_{ab}$, and
our convention on the unit normal to $S$ is that it point 
``radially outward", i.e., into $\Sigma$. 
The surface integral at infinity is simply $4 \pi M $. 
The surface integral at $S$ is just $\kappa A$, where 
$\kappa$ denotes the surface gravity of $\chi^\mu$ on $S$ and $A$ 
is the area of $S$. Therefore, we obtain, 
\begin{equation} 
4 \pi M-\kappa A   =  \int_{\Sigma} \lambda \rho 
\label{Mass4} 
\end{equation} 
Equation (\ref{Mass4}) is our first ``mass formula" for black holes. 
It should be emphasized that this formula applies to an 
arbitrary stationary black hole with a bifurcate Killing horizon, 
with $\rho$ given by eq.(\ref{rho'}), which takes the explicit 
form (\ref{rho}) in the EYM case. 
Both $\rho$ and $\lambda$ are non-negative whenever Einstein's 
equation holds with matter satisfying the strong 
energy condition. 
Hence,  it follows immediately that for all such black holes we have, 
\begin{equation} 
4 \pi M  \geq \kappa A   
\label{Ineq1} 
\end{equation} 
This inequality was recently derived by Visser
\cite{Visser} (using eq.(\ref{Mass1} below)
 for the case of non-rotating black holes.
Our derivation shows that eq.(\ref{Ineq1})
remains be valid for all stationary
black holes, provided only that the matter present in the exterior region
satisfies the strong energy condition.
In particular, if there exist any ``colored excitations" of
the Kerr-Newman black holes (as we conjectured
in \cite{Us}), they must satisfy eq.(\ref{Ineq1}).

To derive our second mass formula, we start with the well known 
integral mass formula of Bardeen, Carter, and 
Hawking \cite{BCH}: 
\begin{equation}
M-\kappa A /4\pi - 2\Omega J_H = 
2 \int_{\Sigma} (T_{\mu \nu} -(1/2) T g_{\mu \nu})
t^\mu n^\nu
\label{Mass1}
\end{equation}
where $T_{\mu \nu}$ is the energy momentum tensor of matter, 
and $J_H$ is the ``angular momentum of the black hole", defined by, 
\begin{equation} 
J_H = (1/16 \pi) \int_S \epsilon_{\mu\nu\sigma\rho} \nabla^\sigma \phi^\rho
\label{Momentun1}
\end{equation}
Equation (\ref{Mass1}) is obtained by starting with the 
Komar formula for the mass 
of a stationary spacetime and converting this surface integral at 
infinity to a volume integral over a hypersurface $\Sigma$ passing 
through $S$ (see, e.g., \cite{WaldB}). It holds for any stationary black 
hole with bifurcate Killing horizon satisfying Einstein's equation with 
arbitrary matter. Note that it is not necessary for the validity of 
 eq.(\ref{Mass1}) 
that $\Sigma$ be a maximal hypersurface. 

We restrict attention, now, to the case where 
the matter is a Yang-Mills field. Then we can write eq.(\ref{Mass1}) 
more explicitly as 
\begin{equation} 
M-\kappa A /4\pi - 2\Omega J_H = (1/4\pi ) \int_{\Sigma} 
 \lambda [ (1/ h)  \eela \Ela + (1/2) \ffl \Fl ] +2 t^a 
 E^b_{\ \Lambda} F_{ab}^{\ \Lambda}/\sqrt h 
\label{Mass2} 
\end{equation} 
A more useful form of eq.(\ref{Mass2}) can be obtained 
by relating $J_H$ to the canonical  angular momentum 
in EYM theory, defined by \cite{Us} 
\begin{equation}
\J_{\infty} = -(1/16 \pi )\oint_{\infty }(2 \phi_b \p +
4\phi^b \Alb \Ela )/\sqrt h dS_a 
\label{(2.40)} 
\end{equation} 
Converting this surface integral to a volume integral 
over $\Sigma$ and using the constraint equations 
as in the derivation of eq.(53) of \cite{Us}, we find, 
\begin{equation}
\J_{\infty} = -
(1/16\pi )\int_{\Sigma } (\p {\mbox{\rm{\pounds}}}_{\phi^i} \h + 
4\Ela {\mbox{\rm{\pounds}}}_{\phi^i} \Ala )/\sqrt h + \J_H  
\label{Jcan} 
\end{equation} 
where $\J_H$ is defined by,
\begin{equation}
\J_H = -(1/16 \pi )\oint_{S}(2 \phi_b \p +
4\phi^b \Alb \Ela )/\sqrt h dS_a 
\label{Jh} 
\end{equation} 
The integral over $\Sigma $ in eq.(\ref{Jcan}) vanishes because 
the axial Killing field $\phi^\mu$ is equal to its tangential 
projection $\phi^i$. 
Thus, we obtain,
\begin{equation} 
 \J_{\infty} = \J_H 
\label{Jcan'} 
\end{equation} 
Furthermore, using the fact that 
\begin{equation} 
 \nabla^\mu \phi^\nu = D^\mu \phi^\nu - 2 \phi_\rho n^{[\mu } K^{\nu] \rho}
\end{equation} 
where $K^{\nu \rho} $ is the extrinsic curvature of $\Sigma $, 
we see that 
the first term in the formula (\ref{Jh}) for $\J_H$ is just $J_H$. 
The second term in eq.(\ref{Jh}) can be computed 
by noting that on $S$, $t^a$ and 
$\phi^a$ coincide (up to the constant $\Omega$) 
as a result of eq.(\ref{(2.45)}) 
and the fact that $\chi^b$ vanishes on $S$. Therefore we can write, 
\begin{equation} 
 4\pi \Omega (J_H -\J_{\infty} ) = -\int_S d S_a t^b \Alb \Ela  /\sqrt h 
\end{equation} 
We now convert this surface integral into a volume integral 
over $\Sigma $, using the the constraint eq.(\ref{(2.11)}).
When we do so, we get no 
contribution from the boundary at infinity on account 
of the asymptotic fall-off 
behavior of $\Alb$ and $ \Ela $. We obtain, 
\begin{equation} 
 4\pi \Omega (J_H -\J_{\infty} ) = \int_{\Sigma}  
( t^b E^a_{\ \Lambda} F_{ab}^{\ \Lambda}/\sqrt h 
+{\mbox{\rm{\pounds}}}_{t^i}( \Ala) E^a_{\ \Lambda}/\sqrt h ) 
\end{equation} 
We now use eq.(\ref{(2.19)}) to substitute for 
${\mbox{\rm{\pounds}}}_{t^i}( \Ala)$ and again use 
the constraint eq.(\ref{(2.11)}). We obtain, 
\begin{equation} 
 4\pi \Omega (J_H -\J_{\infty} ) = -\int_{\Sigma}  
( t^a E^b_{\ \Lambda} F_{ab}^{\ \Lambda}/\sqrt h 
+ \lambda E^a_{\ \Lambda}E_a^{\ \Lambda}/ h ) 
-\int_{\infty} d S_a \lambda E^a_{\ \Lambda}A_0^{\ \Lambda}/\sqrt h 
\end{equation} 
where we used the fact that there is no surface integral 
contribution from $S$ in this equation 
since $\lambda = 0$ on $S$. As shown in \cite{Us}, 
for SU(2)-Yang-Mills theory (or Maxwell theory) the surface integral 
at infinity in this equation yields simply 
$4\pi VQ$, where $V$ is the asymptotic magnitude 
of $A_0^{\ \Lambda}$ and $Q$ is the Yang-Mills electric 
charge at infinity. Therefore, we obtain, 
\begin{equation} 
 4\pi(VQ- \Omega (\J_{\infty} -J_H )) = -\int_{\Sigma}  
( t^a E^b_{\ \Lambda} F_{ab}^{\ \Lambda}/\sqrt h
+ \lambda E^a_{\ \Lambda}E_a^{\ \Lambda}/ h ) 
\label{J3} 
\end{equation} 
Our desired second mass formula is obtained by using this equation 
to eliminate $J_H$ from eq.(\ref{Mass2}). We obtain, 
\begin{equation} 
4 \pi M-\kappa A  + 8\pi (VQ-\Omega \J_{\infty} ) =  \int_{\Sigma} 
 \lambda (  (1/2) \ffl \Fl - (1/ h)  \eela \Ela ) 
\label{Mass3} 
\end{equation} 
Note that eq.(\ref{Mass3}) holds for an arbitrary hypersurface 
$\Sigma$ which is asymptotically orthogonal to $t^\mu$ 
and has $S$ as its inner boundary, i.e., in this formula it is not 
necessary that $\Sigma$ be a maximal hypersurface. 

As previously mentioned, our starting 
point, eq.(\ref{Mass1}), in the derivation 
of eq.(\ref{Mass3}) holds for an arbitrary Einstein-matter system. 
Furthermore, the notion of $\J_{\infty}$ is well defined for any 
Einstein-matter system derivable from a Hamiltonian. 
However, considerable use was made of the 
explicit form of the Yang-Mills field 
equations in deriving 
eq.(\ref{J3}). Thus, it is not clear eq.(\ref{Mass3}) would have 
a close analog for other 
Einstein-matter systems derivable from a Hamiltonian. 

We now subtract eq.(\ref{Mass3}) from eq.(\ref{Mass4}), using 
eq.(\ref{rho}). We obtain, 
\begin{equation} 
 8\pi (\Omega \J_{\infty} -V Q) = 
\int_{\Sigma} \lambda (\P\p + 2 E^a_{\ \Lambda}E_a^{\ \Lambda})/h 
\label{res} 
\end{equation} 
By inspection of eqs.(\ref{res}),
(\ref{Mass4}), and (\ref{rho}), we see that any 
stationary black hole with bifurcate Killing horizon in 
EYM theory satisfies, 
\begin{equation} 
M - \kappa A/4\pi \geq \Omega \J_{\infty} -V Q \geq 0 
\end{equation} 
Furthermore, we obtain directly from eq.(\ref{res}) the following theorem,
which corresponds 
to theorem 3.4 of \cite{Us}:\\

{\bf Theorem 1}:   
A solution of the EYM equations 
describing a stationary black hole with a bifurcate Killing horizon 
that has $\Omega \J_{\infty} -V Q =0$ is static and has vanishing 
electric field on the static slices. 

{\bf Proof}: Since the strong energy condition is satisfied by 
the Yang-Mills field, theorem 4.2 of \cite{WaldC} establishes that 
the exterior region of the black hole can be foliated by maximal 
hypersurfaces with boundary $S$ which are 
asymptotically orthogonal to the timelike 
Killing field $t^\mu$. Applying eq.(\ref{res}) to these
hypersurfaces, we obtain 
$\pi^{ab} =0$ and $ E^a_{\Lambda} = 0. $ 
It then follows directly that $\lambda n^\mu$ is a Killing field
(see \cite{Us}), and, indeed, that $t^\mu = \lambda n^\mu$.\\ 

In our discussion thus far, we have not made use of the first law of 
black hole mechanics for EYM black 
holes (see theorem 2.2 of \cite{Us}), 
which states that the changes in $M$, 
$Q$, $\J_{\infty}$, and $A$ induced 
by an arbitrary asymptotically flat perturbation 
satisfying the linearized EYM equations are related by, 
\begin{equation} 
\delta M + V \delta Q - \Omega \delta \J_{\infty} = 
(1/8 \pi) \kappa \delta A 
\label{Firstlaw} 
\end{equation} 
As emphasized in \cite{Wald}, a formula of this type will exist 
in any Einstein-matter theory having a Hamiltonian formulation. 
For EYM theory, it does not appear possible to derive an integral 
mass formula directly
from eq.(\ref{Firstlaw}). However, in Einstein-Maxwell 
theory, an integral formula can be derived from 
the first law using the additional fact that the 
theory is invariant under the scaling transformation $g_{\mu\nu} 
\rightarrow \alpha^2 g_{\mu\nu}$, 
$A_{\mu} \rightarrow \alpha A_{\mu}$, where $\alpha$ is a 
constant. Under this transformation, 
a solution of the Einstein-Maxwell 
equations is taken into a new solution, 
with $M \rightarrow \alpha M$, 
$V \rightarrow V$, $Q \rightarrow \alpha Q$, 
$\Omega \rightarrow \alpha^{-1}\Omega$, 
$\J_{\infty} \rightarrow \alpha^2\J_{\infty}$, 
$\kappa \rightarrow \alpha^{-1}\kappa$, and 
$A \rightarrow \alpha^2A$. Substituting the linearized perturbation 
associated with this scale transformation into eq.(\ref{Firstlaw}) 
we obtain the mass formula, 
\begin{equation} 
M + VQ - 2\Omega \J_{\infty} = (1/4\pi)\kappa A 
\label{Mass5} 
\end{equation} 
which is valid in the Einstein-Maxwell case. 
Equation (\ref{Mass5}) is equivalent to the ``generalized Smarr 
formula" derived by Carter (see eq.(6.323) of \cite{Carter}
and note that Carter's $\Phi_H$ corresponds to $-V$). 
This mass formula is characterized by the fact that it 
involves only ``surface terms". As our derivation makes clear, a 
similar formula can be obtained for any scale 
invariant Einstein-matter system 
which has a Hamiltonian formulation. 

By combining eqs.(\ref{Mass3}), (\ref{Mass4}), and
(\ref{Mass5}), we can solve for $\Omega \J_{\infty}$
and $VQ$ separately in Einstein-Maxwell theory. We obtain,
\begin{equation} 
\Omega \J_{\infty} = (1/8\pi) 
\int_{\Sigma} \lambda (\P\p /h +  F^{ab} F_{ab}) 
\label{OmegaJ} 
\end{equation} 
and
\begin{equation} 
VQ = - (1/4\pi) 
\int_{\Sigma} \lambda (E^aE_a /h - (1/2)F^{ab} F_{ab}) 
\label{VQ} 
\end{equation} 
The latter equation also can be derived using only Maxwell's 
equations. 

By inspection of eq.(\ref{OmegaJ}), we see that any 
stationary black hole with bifurcate Killing horizon in 
Einstein-Maxwell theory satisfies, 
\begin{equation} 
\Omega \J_{\infty} \geq 0 
\end{equation} 
By the same proof as in theorem 1 above, 
we obtain the following theorem (previously 
proven in the discussion 
following theorem 3.4 of \cite{Us}):\\

{\bf Theorem 2}:   
A solution of the Einstein-Maxwell equation 
describing a stationary black hole with a bifurcate Killing horizon 
that has $\Omega \J_{\infty} =0$ is static and has vanishing 
magnetic field on the static slices. \\ 

Thus, our mass formulas have enabled us to give an elementary 
proof of the staticity theorems of \cite{Us}. \\ 

This research was supported in part by NSF grant PHY-9220644 
to the University of Chicago. 


\eject 
\end{document}